\documentclass[aps,prl,twocolumn,lettersize,showpacs]{revtex4}
\usepackage{epsfig,eepic}
\usepackage{graphicx}
\usepackage[dvips]{color}
\usepackage{amsbsy,latexsym}
\usepackage{amsmath}
\usepackage{amssymb, mathrsfs}
\usepackage[mathscr]{eucal}

\def\d{\operatorname{d}}\def\<{\langle}\def\>{\rangle}
\def\Tr{\operatorname{Tr}}\def\:{\hbox{\bf :}}
\def\vec#1{{\boldsymbol{#1}}}\def\set#1{{\sf #1}}
\def\dag{\dagger}

\def\>{\rangle}
\def\<{\langle}
\def\map#1{\mathcal #1}
\def\spc#1{\mathcal{#1}} 

\def\qed{$\,\blacksquare$\par}

\newtheorem{lemma}{Lemma}

\newtheorem{theorem}{Theorem}

\begin{document}
\title{Optimal cloning of unitary transformations} 
\author{Giulio Chiribella}\email{chiribella@fisicavolta.unipv.it} 
\affiliation{{\em QUIT Group}, Dipartimento di Fisica  ``A. Volta'' and INFN Sezione di Pavia, via Bassi 6, 27100 Pavia, Italy}
\homepage{http://www.qubit.it}
\author{Giacomo Mauro D'Ariano}\email{dariano@unipv.it}
\affiliation{{\em QUIT Group}, Dipartimento di Fisica  ``A. Volta'' and INFN Sezione di Pavia, via Bassi 6, 27100 Pavia, Italy}
\homepage{http://www.qubit.it}
\author{Paolo Perinotti}\email{perinotti@fisicavolta.unipv.it} 
\affiliation{{\em QUIT Group}, Dipartimento di Fisica  ``A. Volta'' and INFN Sezione di Pavia, via Bassi 6, 27100 Pavia, Italy}
\homepage{http://www.qubit.it}
\date{\today}

\begin{abstract}
  After proving a general no-cloning theorem for black boxes, we derive the optimal universal
  cloning of unitary transformations, from one to two copies. The optimal cloner is realized
  by quantum channels with memory, and greately outperforms the optimal measure-and-reprepare
  cloning strategy. Applications are outlined, including two-way quantum cryptographic protocols.

\end{abstract}
\maketitle 

The no-cloning theorem \cite{no-cloning} is one of the cornerstones of Quantum Information, at
the basis of the security of quantum cryptography\cite{ClonRev}, and challenging various protocols,
from optimal estimation to error correction.  Despite the long-dated attention to cloning of quantum
states, cloning of quantum transformations is still a completely unexplored topic.
%Although cloning of quantum
%states has been extensively investigated since the very beginning of
%QIP, cloning of quantum transformations is totally unexplored yet.
Cloning a transformation $\map T$ means exploiting a single use of
$\map T$ inside a quantum circuit, which thus performs the
transformation $\map T \otimes \map T$ on bipartite states.  This
elementary copying task is particularly relevant to the recent trend
in Quantum Information, with the role of the information carrier
played more and more by transformations instead of states, e.~g. in
gate discrimination \cite{dpl,acin}, programming \cite{prog},
teleportation\cite{remcont1,telegate2,telegate3}, and tomography
\cite{dl,debbie}, along with multi-round games \cite{watgut} and
cryptographic protocols like bit commitment with anonymous-state
encoding \cite{protocols1,protocols2}.  Here cloning is the great
absent of the list, partly due to the intrinsic difficulty in treating
manipulations of transformations instead of states. Such a difficulty
has been overcome by the method of Ref.  \cite{QCA}, which
allows one to optimize tasks where the input and output are
transformations.

Cloning quantum transformations can be used for copying quantum software with a limited number of
uses, and in other informational contexts. An interesting application is in the security
analysis of multi-round cryptographic protocols with encoding on secret transformations.  Consider,
for example, the following alternative version of the BB84 cryptographic protocol \cite{bb84}, where
Alice uses two orthogonal bases of unitary transformations instead of states,
${B}_1=\{\sigma_{\mu}\}$ and ${B}_2=\{U \sigma_{\mu}\}$, where $\mu =0,1,2,3$, $\sigma_0$ is the
identity, $\sigma_{1,2,3}$ are the three Pauli matrices, and $U=(I +i \sum_{k=1}^3 \sigma_{k} )/2$
is a rotation of $2\pi/3$ around the axis $\vec n =(1,1,1)/\sqrt{3}$. The protocol works as follows:
Bob prepares an arbitrary maximally entangled state $|B\>$ of two qubits and sends half of it to
Alice, who applies one of the unitaries and sends the output to Bob, as in dense coding \cite{dense}.
Afterwards, Bob measures the two qubits either on the Bell basis $\{(\sigma_{\mu} \otimes I) |B\>\}$
or on the rotated basis $\{(U \sigma_{\mu} \otimes I) |B\>\}$, which are mutually unbiased for
\emph{any} maximally entangled state $|B\>$.  After publicly announcing their choice of bases, and
discarding cases where the bases were different, Alice and Bob use the values of index $\mu$ to
establish a secret key. In this protocol, a na\"ive eavesdropping strategy would be for Eve trying
to estimate the unitaries, by swapping the qubit sent by Bob with a qubit prepared by her---e.g. in  a
known maximally entangled state $|E\>$, half of which she keeps for herself---, and then
intercepting the qubit sent back by Alice. To prevent this attack Alice can randomly ask Bob to send
his half of the entangled state, and to later reveal $|B\>$, so that she can check whether the
received state was actually from Bob. However, Eve can perform coherent attacks that are much more
efficient than na\" ive estimation.  Among coherent attacks, the first and most natural to
investigate is quantum cloning.

Differently from pure states, cloning of transformations is impossible even classically.  This is a
consequence of a \emph{general no-cloning theorem}, holding not only for states, but also for
transformations and any other kind of black boxes (e.g.  measuring devices).  Denoting by $p$ the
minimum of the  worst case error probability in discriminating between two black boxes $\map O_1$ and $\map
O_2$, we have the following theorem, containing as a special case the no-cloning of quantum states
\cite{no-cloning}: \emph{two black boxes cannot be perfectly cloned by a single use unless $p =0$
  (perfect discrimination) or $p=1/2$ (random guess, no discrimination at all)}.  The proof is
simple: If perfect cloning is possible, we can get three copies, perform three times the minimum
error discrimination, and use majority voting to decide the most likely between $\map O_1$ and $\map
O_2$ with worst case error probability $p' = p^2 (3-2 p)$.  Since $p$ is the minimum error
probability, it must be $p \le p'$, whose acceptable solutions are only $p =0$ and $p=1/2$
\cite{1/2}.  Application of the theorem to many black boxes $\{\mathcal O_i\}_{i=1, \dots, k}$
yields the following \emph{cloning-discrimination equivalence}: if for any $i,j$ the error probability
$p_{ij}$ is not $1/2$, then perfect cloning is possible iff perfect discrimination is possible
\cite{proof}.  As a consequence, classical transformations, e.g. permutations of a
classical register, cannot be cloned by a single use (there is no way to discriminate arbitrary
permutations of the letters $\{a,b,c\}$ by evaluation on a single letter).  Likewise, quantum
transformations, e.g.  unitary gates, cannot be cloned by a single use (there is no way to
discriminate arbitrary gates by a single use \cite{dpl}).

%After
%introducing some basics, we explicitly state now a no-cloning theorem
%for algorithms. A (quantum or classical) circuit takes input data into
%output data according to an underlying algorithm, which is described a
%(quantum or classical) channel $\map C$, mapping the input state
%$\rho_{in}$ into the ouput state $\rho_{out} = \map C(\rho_{in})$.
%Cloning the algorithm means exploiting $N$ uses of the channel $\map
%C$ as a subroutine in a computing network that emulates $M$ uses of
%$\map C$, corresponding to the channel $\map C^{\otimes M}$ acting on
%an $M$-partite state. The no-cloning theorem then reads: \emph{two
%  (quantum or classical) channels $\map C_1$ and $\map C_2$ cannot be
%  perfectly cloned unless they can be perfectly distinguished, and
%  viceversa}. Clearly, if two channels can be perfectly copied, then
%they can be perfectly distinguished by tomography. Viceversa, if two
%channels can be perfectly distinguished, then they can be perfectly
%cloned by a classical strategy based on discrimination and subsequent
%repreparation of the corresponding circuit.  

%Equivalently, we can also
%state that two (quantum or classical) channels can be perfectly cloned
%if and only if there is a bipartite input state $\sigma$ such that the
%outputs $\map C_1 \otimes \map I (\sigma)$ and $\map C_2 \otimes \map
%I (\sigma)$ can be perfectly distinguished
%\cite{Hayashi}. %Note that in the quantum case the optimal input state
%$\sigma$ is not always maximally
%entangled \cite{Nota:MaxEnt}.

The existence of a no-cloning theorem immediately rises the question
about the performances of optimal cloners.  In addition to possible
cryptographic applications, the problem has a fundamental interest in itself, as
the relation (if any) between optimal cloning of transformations and cloning
of states is not a priori obvious.

In this Letter we derive the optimal universal cloner, which produces two
approximate copies of a completely unknown unitary gate in dimension
$d<\infty$, showing that entanglement with a quantum memory allows one
to outperform any classical cloning strategy. For qubits  the
global channel fidelity of the clones is $F_{clon}= 46.65\%$,
significantly larger than the fidelity of the optimal
measure-and-prepare scheme $F_{est}= 31 \%$, and of the random guess
(using the given unitary on the first system, and performing a
random  unitary on the second) $F_{ran} = 25 \%$.
Surprisingly, cloning of unitary gates has \emph{no relation} with
cloning of maximally entangled states, in spite of the two sets being commonly considered as equivalent.  Not only cloning maximally entangled
states is always a suboptimal step for cloning unitary gates, but also
any other scheme involving application of the unknown gate to a
maximally entangled state (or any other fixed state) is necessarily
suboptimal.  As it will be shown, this also highlights a fundamental difference
between the two tasks of cloning and learning quantum transformations.
 % This clearly highlightes a
%degradation of the gate as a computational resource.  

%Finally, we
%show that the optimal universal cloning of unitary gates is a stronger
%5task than the universal cloning of pure states: the optimal network derived here a%lso allows to optimally clone pure states.
%\begin{picture}
%\includegraphics{clonunit-2.pstex}
%\end{picture}

\begin{figure}[h]
\begin{center}
%\includegraphics[width=8cm,height=2cm]{figure.eps}\end{center}
%    \scalebox{.25}{\input{clonunit-2.pstex_t}} \qquad
%    \scalebox{.25}{\input{clolearn.pstex_t}}
\setlength{\unitlength}{.28cm}
\begin{picture}(14,5)
  \newsavebox{\isouno} 
  \savebox{\isouno}(2,2)[bl]{
    \multiput(0,0)(2,0){2}{\line(0,1){2}}
    \multiput(0,0)(0,2){2}{\line(1,0){2}}} 
  \newsavebox{\isodue}
  \savebox{\isodue}(2,5)[bl]{ 
    \multiput(0,0)(2,0){2}{\line(0,1){5}}
    \multiput(0,0)(0,5){2}{\line(1,0){2}}}
%%%%% fili %%%%%%
  \put(0,1){\line(1,0){2}}
  \put(4,1){\line(1,0){6}}
  \put(12,1){\line(1,0){2}}
  \put(0,4){\line(1,0){2}}
  \put(4,4){\line(1,0){2}}
  \put(8,4){\line(1,0){2}}
  \put(12,4){\line(1,0){2}}
%%%%% gates %%%%%
  \put(2,0){\usebox{\isodue}}
  \put(10,0){\usebox{\isodue}}
  \put(6,3){\usebox{\isouno}}
%%%%% labels %%%%
  \put(6.6,3.6){$\map U$}
  \put(.2,1.2){$0E$}
  \put(.2,4.2){$0B$}
  \put(4.7,4.2){$1$}
  \put(8.7,4.2){$2$}
  \put(12.3,1.2){$3E$}
  \put(12.3,4.2){$3B$}
  \put(2.4,2.3){$\map A$}
  \put(10.5,2.2){$\map B$}
  \put(6.7,-1.2){a)}
\end{picture}
\quad
\begin{picture}(12,7)
  \newsavebox{\isomono} 
  \savebox{\isomono}(2,2)[bl]{
    \multiput(0,0)(2,0){2}{\line(0,1){2}}
    \multiput(0,0)(0,2){2}{\line(1,0){2}}} 
  \newsavebox{\isoquattro}
  \savebox{\isoquattro}(3,7)[bl]{ 
    \multiput(0,0)(3,0){2}{\line(0,1){7}}
    \multiput(0,0)(0,7){2}{\line(1,0){3}}}
%%%%% fili %%%%%%
  \put(2,1){\line(1,0){6}}
  \put(2,3){\line(1,0){2}}
  \put(6,3){\line(1,0){2}}
  \put(6,4.8){\line(1,0){2}}
  \put(6,6){\line(1,0){2}}
  \put(11,4.8){\line(1,0){2}}
  \put(11,6){\line(1,0){2}}
%%%%% prepa %%%%%
  \put(2,2){\oval(3,4)[l]}
  \put(2,0){\line(0,1){4}}
%%%%% gates %%%%%
  \put(4,2){\usebox{\isomono}}
  \put(8,0){\usebox{\isoquattro}}
%%%%% labels %%%%
  \put(.9,1.9){$\sigma$}
  \put(4.6,2.6){$\map U$}
  \put(0,3.2){0}
  \put(2.7,3.2){1}
  \put(6.3,3.2){$2A$}
  \put(6.3,5){$2E$}
  \put(11.3,5){$3E$}
  \put(6.3,6.2){$2B$}
  \put(11.3,6.2){$3B$}
  \put(8.8,3.3){$\map W$}
  \put(5.7,-1.2){b)}
\end{picture}
\end{center}
\caption{\emph{a) One-to-two cloning of unitaries.}  Two input systems
  are first processed by Eve with channel $\map A$, which entangles
  system 1 and a quantum memory $\spc M$. While the memory $\spc M$ is
  kept by Eve, system 1 is sent to Alice, who applies the secret gate
  $\map U$, and sends back output 2.  Then, Eve applies the channel
  $\map B$, producing two output systems, so that the overall
  transformation from inputs to outputs optimally emulates $\map
  U^{\otimes 2}$. \emph{b) One-to-two quantum learning of unitary.} In
  a training phase, the example $\map U$ is applied locally on the
  bipartite state $\sigma$, and stored in the state
  $\sigma_U=(U\otimes I) \sigma (U^\dag\otimes I)$.  Then, two input
  systems interact with $\sigma_U$, undergoing a transformation that
  optimally emulates $\map U^{\otimes 2}$.}
\label{fig} 
 \end{figure}
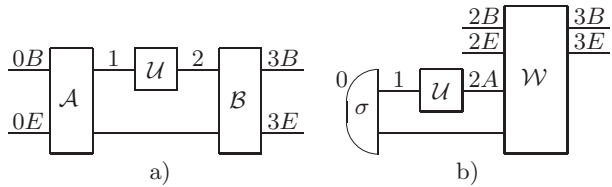

The derivation of the optimal cloner exploits the recent toolbox of \emph{quantum circuit
  architecture theory} \cite{QCA}, which allows optimization of quantum networks for any possible
manipulation of quantum channels, including cloning and estimation.  In this framework any channel
$\map C$ from $\map S (\spc H_{in})$ to $\map S (\spc H_{out})$ [$\map S (\spc H)$ denoting states
on $\spc H$] is described by its Choi operator $C = \map C \otimes \map I ( |I\>\<I|)$, where $|I\>
= \sum_{i=1}^d |i\>|i\>$ is an unnormalized maximally entangled vector in $\spc H_{in} \otimes \spc
H_{in}$. For a unitary channel $\map U (\rho) = U \rho U^\dag$, the Choi operator is $|U\>\<U|$,
with $|U\> =(U\otimes I) |I\>$. A quantum network for $N$-to-$M$ cloning is a network with $N$ open
slots in which the $N$ input copies are inserted, and is also described by a suitable Choi operator
$R^{(N)}$ ($N=1$ for one-to-two cloning, see Fig. \ref{fig}).  If the Hilbert spaces of the inputs
are labeled with even numbers from 0 to $2N$, and the output spaces with odd numbers from 1 to
$2N+1$, the Choi operator $R^{(N)}$ is a non-negative operator on the tensor product
$\bigotimes_{k=0}^{2N+1} \spc H_{k}$ satifsying the recursive normalization condition
\begin{equation}\label{causaln}
%\left\{ 
%\begin{array}{lll}
\Tr_{2N+1}\left[R^{(N)}\right] =I_{2N}\otimes R^{(N-1)},
%\Tr_{2k+1}\left[R^{(k)}\right]&=&I_{2k}\otimes R^{(k-1)} \quad k= 1, \dots, N-1\\
%\Tr_{1} [R^{(0)}]&=&I_0~,
%\end{array} 
%\right.
\end{equation}
where $\Tr_{2N+1}$ denotes the partial trace over the Hilbert space
$\spc H_{2N+1}$ of the $N+1$-th output system, and $R^{(N-1)}$ is the
Choi operator of a network with $N-1$ open slots, which in turn
satisfies Eq. \eqref{causaln} with $N$ replaced by $N-1$. A network
with $N=0$ open slots is a quantum channel from $\map S (\spc H_0)$ to
$\map S(\spc H_1)$, and has the normalization $\Tr_{1} [R^{(0)}] =
I_0$.  Inserting $N$ channels $\map C_1, \dots , \map C_N$ in the $N$
slots of a network, we obtain a new channel $\map C'$ from $\map
S(\spc H_0)$ to $\map S(\spc H_{2N+1})$, with Choi operator given by
\cite{QCA}
\begin{equation}\label{OutputChannel}
C' = \Tr_{1, 2, \dots, 2N} [( I_{0} \otimes C_1^* \otimes \dots \otimes C_N^* \otimes I_{2N+1} ) R^{(N)}]~. 
\end{equation}

In one-to-two cloning ($N=1$) the first input $\spc H_0$ and the last
output $\spc H_3$ must have a bipartite structure, $\spc H_0 = \spc
H_{0B} \otimes \spc H_{0E}$ and $\spc H_{3} = \spc H_{3B} \otimes \spc
H_{3E}$, since the ultimate aim of the network is to mimick the
bipartite channel $\map U_B \otimes \map U_{E}$ on Bob's and
Eve's systems.  Then, the normalization of the Choi operator in
Eq.~\eqref{causaln} gives
\begin{equation}\label{ClonNorm}
\Tr_{3} [R^{(1)}] = I_{2} \otimes R^{(0)}~,\qquad
\Tr_{1}[R^{(0)}]= I_{0}~.
\end{equation} 
Inserting the gate $\map U$ in the network, we obtain the bipartite
channel $\map C'_U$, which according to Eq. (\ref{OutputChannel}) is given by $C'_U= \Tr_{1,2} [(I_{0}
\otimes |U^*\>\< U^*|_{1,2} \otimes I_{3} ) R^{(1)}]$.

We derive now the cloning network for which the channel $\map C'_U$ most
closely resembles $\map U_B \otimes \map U_E$. As a figure of merit we
use the global channel fidelity, uniformly averaged over the unknown
unitaries
\begin{equation}\label{AveF}
\begin{split}
F &= \int \d U ~\frac 1 {d^4} ~   \Tr [ C_U |U\>\<U|^{\otimes 2}]    \\
 & = \frac 1 {d^4} \int \d U~ \<U| \<U|\<U^*| R^{(1)} |U\>|U\>|U^*\>~.
\end{split}
\end{equation} 
%(here we dropped out the numeration of Hilbert spaces, to avoid
%overload of notation). 
Note that $F=1$ if and only if $\map C_U = \map U^{\otimes 2}$ for any $U$, corresponding to perfect
cloning. Exploiting symmetry then provides a radical simplification of the problem:
\begin{lemma}\label{Lemma}
  The optimal cloning network maximizing the channel fidelity
  (\ref{AveF}) can be assumed without loss of generality to be
  \emph{covariant}, i.e. with a Choi operator $R^{(1)}$ satisfiying
  the commutation relation
\begin{equation}
[R^{(1)}, V_0^{\otimes 2} \otimes V_1^* \otimes W_2^*\otimes W_3^{\otimes 2}]=0 \quad \forall V,W \in \mathbb {SU} (d)~. \label{CovR}
\end{equation}  
\end{lemma}
{\bf Proof.} Let $R^{(1)}$ be optimal. Then take its average
$\overline{R^{(1)}} = \int \d V\int \d W~ \map V_0^{\otimes 2} \otimes
\map V_1^* \otimes \map W_2^* \otimes \map W_3^{\otimes 2} (R^{(1)})$, where
$\map V, \map V^*,\map W^*, \map W$ are the unitary channels
corresponding to $V, V^*, W^*, W$.  It is immediate to see that
$\overline {R^{(1)}}$ statisfies Eqs. (\ref{CovR}) and
(\ref{ClonNorm}), and has the same fidelity as $R^{(1)}$. \qed The
representation $V^{\otimes 2} \otimes V^*$ in Eq.  (\ref{CovR}) can be
decomposed into irreducible blocks as follows.  First, one has
$V^{\otimes 2} = V_+ \oplus V_-$, where $V_{\pm}$ is the irreducible
block acting in symmetric (antisimmetric) subspace $\spc H_{\pm}
\subset \spc H^{\otimes 2}$, of dimension $d_{\pm}= d(d\pm1)/2$.
Then, one can further decompose $V_+\otimes V^*= V_{\alpha,+} \oplus
V_{\beta,+}$, where $V_{\alpha,+} (V_{\beta,+})$ is the irreducible
block acting on the subspace $\spc H_{\alpha,+} (\spc H_{\beta,+})
\subset \spc H_+ \otimes \spc H$, of dimension $d_{\alpha} =d$
($d_{\beta} = d(d_+-1)$).  Similarly,  $V_- \otimes V^*= V_{\alpha,-} \oplus V_{\gamma,-}$, corresponding
to irreducible subspaces $\spc H_{\alpha,-} , \spc H_{\gamma,-}
\subseteq \spc H_-\otimes \spc H$ of dimensions $d_{\alpha}=d,
d_{\gamma}= d(d_- -1)$, respectively. Note that the subspaces $\spc
H_{\alpha,+}$ and $\spc H_{\alpha,-}$ carry equivalent
representations, and that for qubits the block $\spc H_{\gamma,-}$ does not show up. By Schur lemmas, the Choi operator $R^{(1)}$ in Eq.
(\ref{CovR}) must be of the form $
%\begin{equation}
R^{(1)} = \sum_{\mu,\nu \in \set S} ~\sum_{i,j,k,l=\pm}~ r^{\mu\nu}_{ik,jl}~ T^{\mu}_{ij} \otimes T^{\nu}_{kl} 
%\end{equation}
$
where $\set S=\{\alpha,\beta,\gamma\}$, $r^{\mu \nu}_{ik,jl}$ is a
non-negative matrix for any $\mu, \nu$, and $T^{\mu}_{ij}=
\sum_{n=1}^{d_{\mu}} |\mu,i,n\>\<\mu,j,n|$ is the isomorphism between
the two equivalent subspaces $\spc H_{\mu,i}$ and $\spc H_{\mu,j}$
($T^{\beta}_{+-}=T^{\beta}_{-+}= T^{\gamma}_{+-}=T^{\gamma}_{-+}=0$).
%Recall that $T^{\mu}_{ij}$ and $T^{\nu}_{kl}$ act on $\spc H_0 \otimes
%\spc H_1$ and $\spc H_2 \otimes \spc H_3$, respectively. 
Exploiting this fact, we obtain for the fidelity the following
expression
\begin{equation}
F= \frac 1 {d^4} \sum_{\mu \in \set S} \sum_{i,j=\pm} d_{\mu}  r^{\mu\mu}_{ii,jj}
\end{equation}
while the normalization constraint of Eq. (\ref{ClonNorm}) becomes
\begin{equation}\label{tmu}
\sum_{\mu, \nu}~  d_{\mu} d_{\nu}  t^{\mu\nu}_{i} = d_i d~ \qquad  t^{\mu\nu}_{i}:= \sum_k r^{\mu\nu}_{ik,ik},~  i=\pm~. 
\end{equation}
We are now ready to derive the optimal cloner:
\begin{theorem}\label{theo}
  The fidelity of the optimal universal cloning of unitary transformations is $F_{clon}= (d+ \sqrt
  {d^2-1})/d^3$. The value $F_{clon}$ is achieved by a network as in Fig. \ref{fig}a
  with pre-processing channel $\map A$ from $\map S(\spc H_{0B}\otimes
  \spc H_{0E})$ to $\map S(\spc H_{0A} \otimes \spc M)$,  $\spc M$ being
  a memory qubit, given by
\begin{equation}
\map A (\rho) = \sum_{i,j =\pm}  \Tr_{0E}[P_i \rho P_j] \otimes |i\>\<j|
\end{equation}
($P_{\pm}$ orthogonal projector on $\spc H_{\pm}$, and $\{|+\>,\|-\>\}$ orthonormal basis for $\spc M$), and
post-processing channel $\map B$ from $\map S(\spc H_{0B} \otimes \spc
M)$ to $\map S(\spc H_{3B} \otimes \spc H_{3E})$, given by
\begin{equation}
\map B(\sigma) = \sum_{i,j=\pm}   \frac d {\sqrt{d_i d_j}} ~ P_i \left[\<i| \sigma |j\> \otimes I_{3E}   \right] P_j~.
\end{equation}
Accordingly, the approximate cloning of $U$ is a channel $\map C'_U$
from $\map S(\spc H_{0B} \otimes \spc H_{0E})$ to $\map S(\spc H_{3B}
\otimes \spc H_{3E})$:
\begin{equation}\label{CU}
\begin{split}
\map C'_U (\rho) &= \map B \circ (\map U \otimes \map I_{\spc M}) \circ \map A (\rho)\\
 &=\sum_{i,j=\pm}  \frac d {\sqrt{d_i d_j}} ~ P_i\left[ U \Tr_{0E} [P_i \rho P_j] U^{\dag}  \otimes I  \right]P_j~.
\end{split}
\end{equation}
\end{theorem}
{\bf Proof.} For the fidelity we have the following bound: 
\begin{equation*}
\begin{split}
F &\le \frac 1 {d^4} \left( \sum_{i=\pm} \sqrt{\sum_{\mu \in \set S} d_{\mu}  r^{\mu\mu}_{ii,ii}} \right)^2  \le  \frac 1 {d^4}\left( \sum_{i=\pm} \sqrt{\sum_{\mu \in \set S} d_{\mu}  t^{\mu\mu}_i} \right)^2\\
&\le \frac 1 {d^4} \left( \sum_{i=\pm} \max_{\mu \in \set S}  \left\{\sqrt{\frac{d_i d} {d_{\mu}}} \right\} \right)^2~.
\end{split}
\end{equation*}
The first inequality comes from Schwartz inequality applied to the
non-negative matrix $a_{i,j}=\sum_{\mu} d_{\mu} r^{\mu\mu}_{ii,jj}$, the
second from the definition of $t_i^{\mu\mu}$ (Eq. (\ref{tmu})), and
the third from  constraint  (\ref{tmu}).  Since the maximum in the bound is
achieved for minimum $d_{\mu}$, i.e. for $\mu=\alpha$, we have $F\le
1/d^4 (\sqrt {d_+} + \sqrt{d_-})^2 \equiv
F_{clon}$. % It is easy to see that the Choi operator
%$R^{(1)} = \sum_{i,j=\pm} \sqrt{d_i d_j} (T^{\alpha}_{ij}\otimes
%T^{\alpha}_{ij})/d$ is normalized and achieves the bound.  
To conclude achievability, we directly compute the fidelity between $\map C'_U$ (Eq. (\ref{CU})) and $\map U^{\otimes 2}$, which yields $F(\map C'_U, \map U^{\otimes 2}) =F_{clon} \ \forall U$. \qed

Let us now clearify the meaning of the pre- and post-processing
channels $\map A$ and $\map B$ in the optimal network. First, channel
$\map A$ can be extended to a unitary interaction between the input
systems $\spc H_{0B}, \spc H_{0E}$ and the memory $\spc M$: $\map A
(\rho) = \Tr_{0E} [V (\rho \otimes |0\>\<0| ) V^\dag ]$, where $|0\> =
(|+\> +|-\>)/\sqrt{2} \in \spc M$, and $V$ is the controlled-swap $V=
I \otimes |+\>\<+| + S \otimes |-\>\<-|$, $S |\phi\>|\psi\>=
|\psi\>|\phi\>$.  Such an extension has a very intuitive meaning in
terms of quantum parallelism: for bipartite input $|\Psi\>_{BE}$ the
single-system unitary $U$ is made to work on both $B$ and $E$ by
applying it to the superposition $|\Psi\>_{BE}+S|\Psi\>_{BE}$ and
discarding $E$.  Less intuitive, and much more intriguing, is the
meaning of channel $\map B$. It is an \emph{extension of optimal universal
  cloning} of pure states \cite{OptClon}: if system $\spc H_{0B}$ and
the ancilla $\spc M$ are prepared in the state $|\psi\>|+\>$, then we
obtain $\map B (|\psi\>\<\psi| \otimes |+\>\<+|) = d/d_+ \left[P_+
  (|\psi\>\<\psi| \otimes I) P_+\right]$, which are indeed two optimal
clones of $|\psi\>$.  This means that realizing the optimal cloning of
unitaries is a harder task than realizing the optimal cloning of
states: an eavesdropper that is able to optimally clone unitaries must
also be able to optimally clone pure states. This suggests that
cryptographic protocols based on gates (such as the two-way protocol
in the introduction) might be harder to attack than protocols based on
states.

The performances of the optimal cloner crucially depend on
entanglement with the quantum memory $\spc M$: Suppose that after
channel $\map A$ the ancillary qubit $\spc M$ decoheres on the basis
$\{|+\>,|-\>\}$. Then, the approximate cloning of $U$ is no longer
given by Eq.  (\ref{CU}), but rather by its decohered version
$\widetilde{\map C'}_U(\rho) = \sum_{i=\pm} d/d_i ~ \left[ P_i ( U
  \Tr_{0E} [P_i \rho P_i] \otimes I) P_i \right]$.  Direct calculation
of the fidelity in this case gives $F_{deco} = 1/d^2$, which is
exactly the same fidelity $F_{ran}$ that one would achieve by applying $U$ on the first system and by performing a
randomly chosen unitary on the second.  For large $d$, the optimal
fidelity achieved with the quantum memory is essentially twice this
value.  Another classical cloning strategy  would be to optimally
estimate the unknown unitary, getting an estimate $\hat U$, and then
performing $\hat U^{\otimes 2}$ on the input systems $\spc H_{0B}$ and
$\spc H_{0E}$. Using the optimal estimation strategy of Ref.
\cite{EntEstimation} we can readily evaluate the fidelity of estimation to be $F_{est}=6/d^4$ for $d>2$, $F_{est}= 5/16$ for $d=2$.
Note that, as far as it concerns the \emph{global fidelity}, for $d>
2$  estimation is by far worse than the
crude decohered strategy described above.

We answer now a natural question: since there is a canonical isomorphism between unitaries $U$ and
maximally entangled states $|U\>=(U\otimes I) |I\>$, one might wonder whether the optimal cloning of
$U$ can be achieved via cloning of the state $|U\>$. Surprisingly at first sight, the answer is
negative. In order to prove this fact, we put ourselves in a slightly more general scenario: we
apply the unknown gate $U$ to an arbitrary bipartite state $\sigma \in \map S (\spc P)$ (not
necessarily maximally entangled), and use the state $\sigma_U:= (U\otimes I)\sigma (U^\dag \otimes
I)$ to program a transformation $\map L_U$ on the two systems $\spc H_{0B}, \spc H_{0E}$, given by
$\map L_U (\rho)= \Tr_{\spc P}[W (\rho \otimes \sigma_U) W^\dag]$, where $W$ is a suitable
interaction.
%\begin{figure}
%\end{figure}
Again, the goal is to maximize the fidelity between $\map L_U$ and $\map U_B \otimes \map U_E$. This
is an elementary instance of \emph{quantum learning}, in which a training set of examples---$N$ uses
of the unknown $\map U$---is provided in a first stage ($N=1$ here), and, after the training has been
concluded, the learning machine is asked to optimally emulate $\map U^{\otimes M}$ ($M=2$ here).  Using the same method illustrated for cloning, we can find the optimal
one-to-two learning network with Choi operator $L^{(1)}$, for which the normalization (\ref{ClonNorm})
now reads
\begin{equation}\label{NormL}\Tr_{3} [
L^{(1)}] = I_2 \otimes \rho_1~, \qquad \Tr[\rho_1]=1~,
\end{equation} 
where $I_2$ acts on the tripartite space $\spc H_2= \spc H_{2B}\otimes
\spc H_{2E} \otimes \spc H_{2 A}$ and the space $\spc H_0$ is
one-dimensional (see Fig. \ref{fig}).  Using the symmetry argument of
Lemma \ref{Lemma}, we can restrict the optimization to Choi operators
satisfying $[L^{(1)}, V_1^* \otimes V^{\otimes 2}_{2B, 2E} \otimes
W_{2A}^* \otimes W_{3B,3E}^{\otimes 2}]=0$, i.e. of the form $L^{(2)}
= \sum_{\mu,\nu \in \set S} \sum_{i,j,k,l=\pm} l^{\mu\nu}_{ik,jl}
T_{ij}^{\mu} \otimes T^{\nu}_{kl}$. The normalization of Eq.
(\ref{NormL}) then becomes $\sum_{\nu, m} d_{\nu} l^{\mu\nu}_{im,jm}=
\delta_{ij}$ for any $\mu \in \set S,$ and $i,j=\pm$. Maximizing the
fidelity under this constraint we then obtain the maximum value
$F_{learn} = 6/d^4$ for $d>2$ and $F_{learn}=5/d^4$ for $d=2$, exactly
the same value of optimal estimation.  Therefore, \emph{any} scheme
based on the application of $U$ on a fixed input state will be
extremely poor compared to the optimal cloner. This highlights the
fundamental difference between quantum learning and cloning: in
learning one has to first apply the unknown gate $U$ to a fixed state
$\sigma$, which implies an irreversible degradation of its
computational power. Note that the difference between cloning and
learning is a specific treat of quantum channels, while for states
there is no difference between the two tasks. Moreover, one can regard
the quantum learning of unitary transformations as a special case of
gate programming \cite{prog}, e.~g. for one-to-two learning the
program is of the form $\sigma_U= (U\otimes I)\sigma (U^\dag
\otimes I)$ and the target is $U\otimes U$.

%In conclusion, we addressed the problem of optimal channel cloning and
%presented the optimal network emulating two uses of an unknown unitary
%from a single use. We showed that the key feature allowing the optimal
%network to achieve better performances than any classical cloning is
%entanglement with a quantum memory. In addition, we showed that the
%optimal cloning network contains a post-processing interaction which
%extends the optimal cloning of pure states, implying that gate cloning
%is a harder kind of information processing than pure state cloning,
%with interesting consequences for cryptographic protocols based on
%gates instead of states.  Finally, we highlighted the fundamental
%difference between quantum cloning and quantum learning, proving as a
%byproduct that optimal cloning of unitaries cannot be achieved via
%optimal cloning of maximally entangled states.

In conclusion, in this Letter we proved a general no-cloning theorem for black boxes,
and derived the optimal universal cloning of unitary transformations
from one to two copies. The optimal cloner is realized via a quantum
channel with memory, and greately outperforms the optimal
measure-and-prepare strategy.  Exploring the deep relations among
cloning, learning, and programming of quantum transformations is a natural development of our work and an
interesting avenue for future research.

\par {\em Acknowledgments.---} We thank S. De Zordo for preliminary calculations in his master
thesis. GC is grateful to M. Murao, M.  Hayashi, and A. Winter for useful and enjoyable discussions.
This work is supported by the EC through the networks SECOCQ and CORNER.


\begin{thebibliography}{99}
\bibitem{1/2} Note that $p$ cannot exceed $1/2$: with a random guess one can always reduce the error probability to $1/2$.
\bibitem{proof} If perfect discrimination is possible, discrimination
  plus repreparation gives perfect cloning. Viceversa, if perfect
  cloning is possible, the no-cloning theorem implies $p_{ij}=0$ for
  any $i \not=j$. Then we need only $k-1$ perfect clones to achieve
  perfect discrimination via $k-1$ pairwise tests.
\end{thebibliography}
\end{document}